\documentclass{INTERSPEECH2023}

\usepackage{multirow,cite} 
\usepackage{comment} 

\interspeechcameraready

\title{Improving the Gap in Visual Speech Recognition \\
Between Normal and Silent Speech Based on Metric Learning}
\name{Sara Kashiwagi$^{1 *}$, Keitaro Tanaka$^{1 *}$, Qi Feng$^1$, Shigeo Morishima$^2$}
\address{
  $^1$Waseda University, Japan\\
  $^2$Waseda Research Institute for Science and Engineering, Japan}
\email{sara.kashiwagi@moegi.waseda.jp, 
phys.keitaro1227@ruri.waseda.jp, 
fengqi@ruri.waseda.jp, 
shigeo@waseda.jp}

\begin{document}

\maketitle
 
\begin{abstract}
This paper presents a novel metric learning approach to address the performance gap between normal and silent speech in visual speech recognition (VSR). The difference in lip movements between the two poses a challenge for existing VSR models, which exhibit degraded accuracy when applied to silent speech. To solve this issue and tackle the scarcity of training data for silent speech, we propose to leverage the shared literal content between normal and silent speech and present a metric learning approach based on visemes. Specifically, we aim to map the input of two speech types close to each other in a latent space if they have similar viseme representations. By minimizing the Kullback-Leibler divergence of the predicted viseme probability distributions between and within the two speech types, our model effectively learns and predicts viseme identities. Our evaluation demonstrates that our method improves the accuracy of silent VSR, even when limited training data is available.
\end{abstract}

\renewcommand{\thefootnote}{\fnsymbol{footnote}}
\footnote[0]{$^*$Equal contribution.}

\vspace{-0.5mm}
\section{Introduction}
Visual speech recognition (VSR), also known as lipreading, aims to recognize spoken words based solely on visual interpretation of lip movements~\cite{Chung2016LipRS, Ma2022VisualSR}. Since VSR is independent of background noise or sound quality, it has attracted significant attention as an alternative to traditional speech recognition, which relies on audio input. VSR also serves as a valuable communication tool for individuals with hearing or speech impairments. Consequently, VSR has already been implemented in practical systems such as the silent speech interface (SSI)~\cite{Sun2018LipInteractIM, Pandey2021LipTypeAS}, which enables users to input text or operate smartphones through non-vocalized (i.e., silent) speech.


One of the major challenges in VSR is the difference in speech format, especially between vocalized (i.e., normal) and silent speech. Studies indicate that in silent speech, individuals tend to move their mouths more extensively than they do in normal speech~\cite{Janke2010ImpactOL, Bicevskis2016EFFECTSOM}. Currently, most VSR datasets contain normal speech data, as they are readily obtainable from TV programs~\cite{Afouras2018DeepAS} or the internet~\cite{Afouras2018LRS3TEDAL}. Consequently, VSR models intended for silent speech are typically trained on normal speech data, resulting in sub-optimal performance~\cite{Florescu2010SilentVV, Petridis2018VisualOnlyRO}.


\begin{figure}[t]
  \centering
  \includegraphics[width=0.9\linewidth]{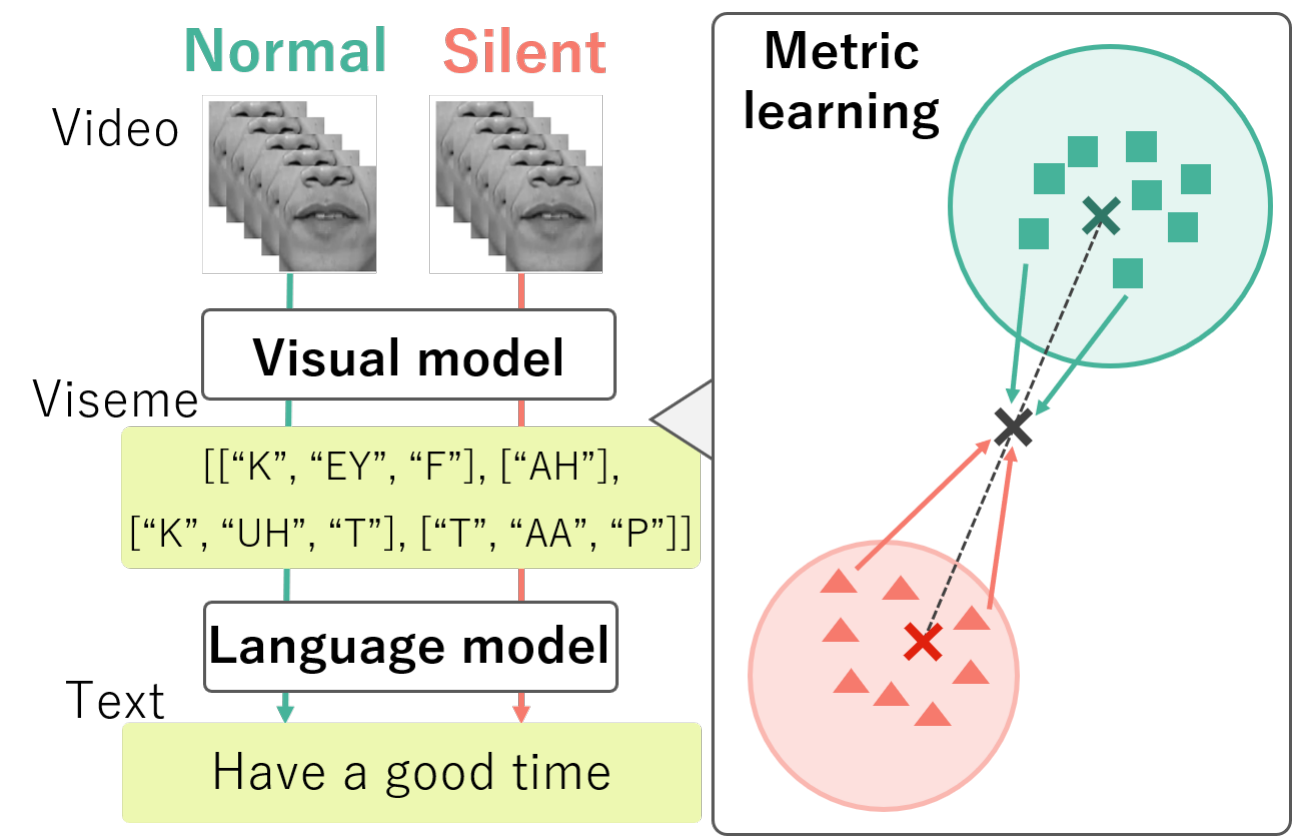}
  \vspace{-1.5mm}
  \caption{
  Our method employs metric learning in a latent space to establish relationships between normal and silent speech instances through shared visemes.
  }
  \vspace{-2mm}
  \label{fig:overview}
  \vspace{-2.5mm}
\end{figure}


A common approach to address the issue of data variance in VSR is to augment the training data to cover all possible variants by collecting or creating them automatically. This method has proven successful in speaker-independent~\cite{Chung2016LipRI} and pose-invariant lipreading~\cite{Anina2015OuluVS2AM, Chung2017LipRI}. However, it is challenging to apply this approach to silent speech recognition due to the scarcity of public datasets and the need for manual data collection. In practice, the largest available dataset for normal speech contains over 150,000 samples~\cite{Afouras2018LRS3TEDAL}, while the largest dataset for silent speech has only up to 4,600 samples~\cite{Petridis2018VisualOnlyRO}.

In this paper, we present a novel metric learning method based on visemes. We observe that normal and silent speech may visually differ, but share the same literal contents (i.e., underlying texts). In the context of VSR for silent speech, it is desirable for a model to recognize input videos from normal and silent speech as similar when they have the same literal representation (or expected outputs). Visemes, the smallest units of visually indistinguishable lip movements~\cite{Lee2002AudiotoVisualCU} that are uniquely derived from phonemes and words, can capture the shared information between normal and silent speech\footnote[2]{This is because visemes are uniquely decided based on a group of phonemes, the smallest units of speech sounds~\cite{Cappelletta2011VisemeDC}, which are also uniquely derived from words or sentences.}. Thus, we propose to focus on visemes to develop our novel metric learning method.

To facilitate the proposed learning method and address the accuracy imbalance between normal and silent speech, 
we propose to utilize a two-stage VSR model~\cite{Fenghour2020LipRS, Peymanfard2021LipRU}, comprising a video-to-viseme (i.e., visual) model and a viseme-to-text (i.e., language) model. Our method enforces the visual model to map input videos of normal and silent speech close to each other in a latent space, provided they share the same viseme (see Fig.~\ref{fig:overview}). Specifically, each point in the latent space represents a probability distribution as estimated visemes, and we minimize the Kullback-Leibler (KL) divergence between the two probability distributions of the two types of speech. Additionally, our method extends to visemes within the same speech type, but presenting in different positions within the same text. This allows the model to learn viseme identities, as well as to estimate visemes independently.

\begin{figure*}[t]
   \centering  \includegraphics[width=0.8\linewidth]{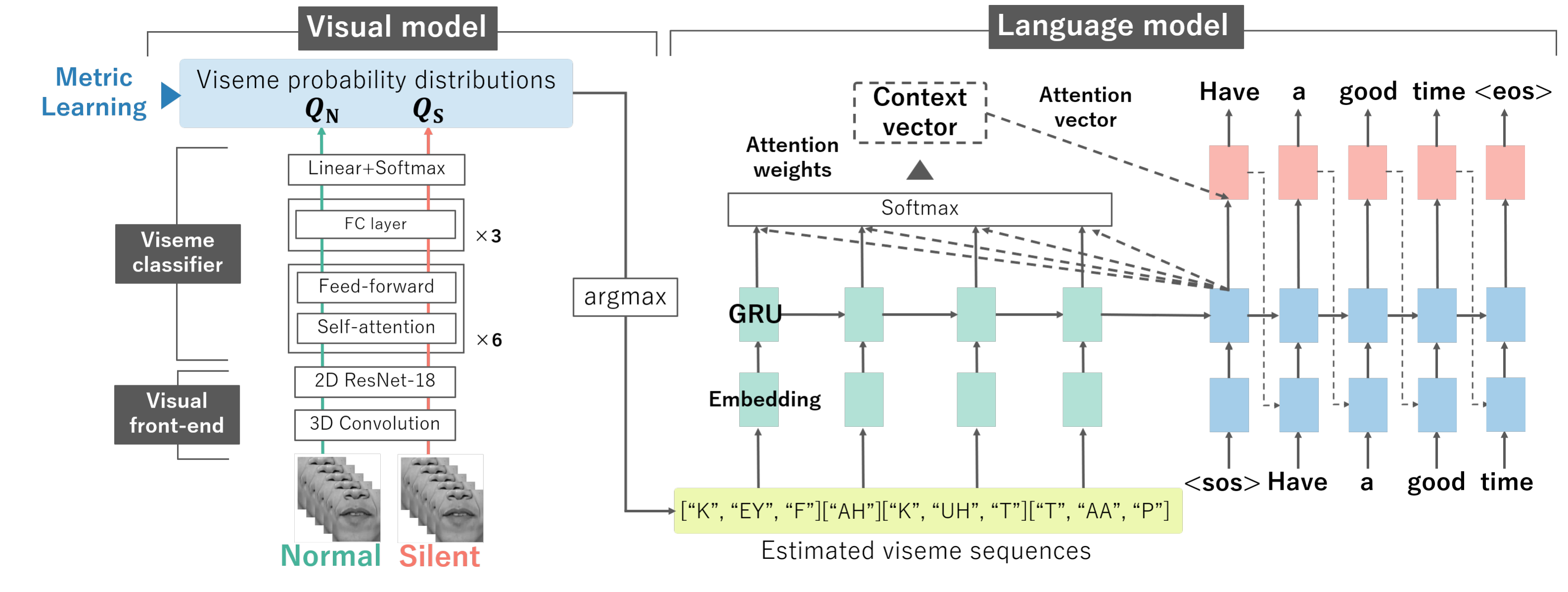}
  \vspace{-3.5mm}
  \caption{
  Two-stage VSR model incorporating a visual model and a language model. 
  Our approach incorporates metric learning into the visual model outputs, representing viseme probability distributions of normal and silent speech data inputs.
  }
  \vspace{-4.5mm}
   \label{fig:framework}
 \end{figure*}

The main contribution of this study is to propose a novel learning method to leverage already existing yet limited amounts of silent speech data more efficiently. Our method achieves better performance without collecting single additional speech data, even that of normal speech. In addition, we demonstrate that our method's effectiveness is further enhanced by accounting for data imbalance between the two speech types. Experimental results show that our method improves the performance of the model by nearly 3\% from a baseline. Furthermore, the model trained with 400 silent speech data using our method performs comparably to the baseline model trained with twice the amount of data.


\section{Related work}
This section provides a summary of the existing literature regarding the differences in lip movements between normal and silent speech. 
Additionally, we explore the advantages of incorporating visemes as an intermediate feature in lipreading models as compared to alternative approaches.

\subsection{Normal speech and silent speech}
Research examining lip movements during normal and silent speech has utilized facial electromyography (EMG) signals~\cite{Janke2010ImpactOL} and ultrasound images~\cite{Bicevskis2016EFFECTSOM} to establish significant differences between the two types of speech. 
Specifically, studies have shown that the absence of auditory feedback in silent speech leads speakers to rely more on somatosensory feedback to ensure proper pronunciation, resulting in larger lip movements, particularly when articulating bilabial consonants such as ``p'', ``b'', and ``m''. 
Notably, two additional studies~\cite{Florescu2010SilentVV, Petridis2018VisualOnlyRO} have reported a decline in performance during silent speech recognition, a challenge that remains unresolved.
Petridis et al.~\cite{Petridis2018VisualOnlyRO} observed that a model trained on normal speech achieved 8.5\% lower accuracy when tested on silent speech compared to normal speech using their original dataset.


\subsection{Viseme-level lipreading}
Current research on VSR has primarily focused on developing end-to-end models~\cite{Ma2022VisualSR, Afouras2018DeepAS}, which necessitate significant amounts of paired video and text data to capture the context of the inputted utterances.
Alternatively, lipreading models that generate other tokens, such as sub-words~\cite{KR2021SubwordLL}, phonemes~\cite{Shillingford2018LargeScaleVS, Bear2019AlternativeVU}, or visemes~\cite{Fenghour2020LipRS, Peymanfard2021LipRU}, have also been investigated. 
These models employ an external language model that predicts spoken texts based on the output tokens,
facilitating the prediction of unseen words during training. 
Moreover, several studies~\cite{Fenghour2020LipRS, Peymanfard2021LipRU} have identified a common challenge in both end-to-end and viseme-level two-stage models, namely, differentiating homophones: words that possess unique meanings but similar lip movements. This issue arises due to the fact that the number of visemes is significantly smaller than the number of phonemes. 
In the latter models, language models 
can have decoding systems that correspond to homophones, 
leading to more efficient utilization of visual information 
from small datasets.


\section{Proposed method}
We first explain the model architecture
 and then introduce two kinds of metric learning,
 inter-speech and within-speech types,
 applied to the outputs of the visual model
 (Fig.~\ref{fig:framework}).


\subsection{Two-stage VSR model}
As a pioneering work for mitigating the imbalance between normal and silent speech,
 we aim to explore the potential of viseme information.
In our work, we develop a two-stage model for VSR
 that explicitly incorporates this information following Fenghour et al.~\cite{Fenghour2020LipRS, Fenghour2019DecoderEncoderLF, Fenghour2021AnEC}.
It consists of a visual model 
 that predicts from input videos viseme sequences
 and a language model 
 that generates from the predicted viseme sequences the output texts.
Note that the input video is a sequence of colored images.
We convert them to grayscale and crop around their lips 
 using Dlib~\cite{Kazemi2014OneMF}.
It turns into a tensor with the shape of $T \times 112 \times 112$,
 where $T$ is the number of frames.


The visual model consists of a visual front-end
 that extracts spatial and temporal features from the input tensor
 and a viseme classifier
 that outputs the predicted viseme sequences.
In the visual front-end,
 a 3D convolutional layer captures the changes in lip shapes,
 and then 2D ResNet-18 follows
 to output a 512-dimensional feature vector for each video frame.
For the viseme classifier, we employ an encoder-decoder architecture.
Specifically,
 we utilize a transformer's encoder 
 with repeated self-attention and feed-forward networks
 and three fully connected layer blocks 
 with 1,024, 2,048, and 1,024 nodes, respectively.
The last layer of the classifier 
 consists of a linear layer and a softmax function,
 which yields probability distributions for visemes.
Each frame is classified into 17 classes, 
 including 13 viseme classes 
 and four special characters for space, start of sentence (SoS), end of sentence (EoS), and padding. 

The language model is based on an attention architecture and
 two layers of the gated recurrent unit (GRU) with 256 nodes.
Specifically,
 the first layer of the GRUs serves as an encoder,
 where attention weights are calculated against its output,
 and the second one does as a decoder.
The language model takes as inputs the viseme sequences predicted by the visual model,
 where each combination of visemes is regarded as a distinct class,
 and every possible word is considered a target text label.
During the training of the language model,
 the decoder uses the ground truth as input.
In contrast, during inference,
 the decoder uses a context vector calculated from attention weights that access the encoder's last hidden state to predict the next word based on all previously predicted words.
The visual model and the language models are trained
 so that cross-entropies between their output and the ground truth
 are minimized.
We denote the cross-entropy loss function for normal speech data
 as $\mathcal{L}_\mathrm{NCE}$
 and that for silent speech data
 as $\mathcal{L}_\mathrm{SCE}$.

\subsection{Metric learning for inter-speech type}
Although normal and silent speech exhibits different lip movements, 
 the predicted viseme probability distributions should be similar  
 regardless of speech type,
 as long as they have the same viseme. 
To fill this gap, 
 we propose two loss functions based on KL divergence, 
 derived from different handling manners of the two types of speech.
Let $P_\mathrm{N}$ ($P_\mathrm{S}$) be the target distribution 
 of normal (silent) speech 
 and $Q_\mathrm{N}$ ($Q_\mathrm{S}$) be the predicted distribution 
 of normal (silent) speech. 
For both speech types, the predicted distribution 
 $Q = \{\bm{q}_{1}, ..., \bm{q}_{L}\} \in \mathbb{R}^{C \times L}$ 
 is generated by the visual model, 
 where $L$ is the length of the viseme sequence 
 and $C$ is the number of viseme classes.
The ground truth of the viseme sequence corresponding to $Q$ is given as 
 $Y = \{y_{1}, ..., y_{L}\} \in \mathbb{N}^{L}$, 
 where ${(y_{l})}^{L}_{l=1} \in \{{1, ..., C}\}$.
We use the average of all distributions classified into the same viseme within each speech type as the representative distribution for the viseme label.
The alignment of distributions is represented as $S = \{\bm{s}_{1}, ..., \bm{s}_{C}\} \in \mathbb{R}^{C \times C}$, where ${(\bm{s}_{c})}^{C}_{c=1} \in \mathbb{R}^{C}$ is the representative distribution for viseme class $c$ defined as:
\begin{align}
    \bm{s}_{c} = \dfrac{\sum_{y_{l} = c} \bm{q}_{l}}{m},
    \label{equation:eq3}
\end{align}
where $m$ represents the number of times the viseme label $c$ appears in the true viseme sequence $Y$, i.e., $m = \#(y_{l} = c)$.
By aligning each distribution included in $S$ with the viseme label of $Y$, we create the target distribution $P = \{\bm{p}_{1}, ..., \bm{p}_{L}\} \in \mathbb{R}^{C \times L}$ (for example, if $y_{2} = 8$, $\bm{s}_{8}$ is allocated to $\bm{p}_{2}$).

\vspace{-0.5mm}
\subsubsection{Metric learning with single KL divergence}
\vspace{-1mm}
In this method, we regard normal speech as the primary speech type
 and silent speech as its variant.
Specifically, we leverage the KL divergence to bring 
 the predicted distribution of silent speech 
 close to the target distribution of normal speech 
 by minimizing the following $\mathcal{L}_\mathrm{KL}$:
\begin{align}
    \mathcal{L}_\mathrm{KL} &= \mathcal{D}_\mathrm{KL}(P_\mathrm{N}||Q_\mathrm{S}) = {P_\mathrm{N}}\log\dfrac{P_\mathrm{N}}{Q_\mathrm{S}}.
    \label{equation:eq4}
\end{align}

\subsubsection{Metric learning with weighted KL divergence}
\vspace{-1mm}
Next, we regard both types of speech as variants of speech in general
 (i.e., we regard them as of equal importance).
Such a point of view lets us consider a weighted distribution
 that depends on the number of training data used for each speech type
 to the target distribution of the KL divergence.
The weighted distribution $M$ can be defined as:
\begin{align}
    M = \frac{{N}_\mathrm{N}P_\mathrm{N} + {N}_\mathrm{S}P_\mathrm{S}}{{N}_\mathrm{N}+{N}_\mathrm{S}},
    \label{equation:eq5}
\end{align}
where ${N}_\mathrm{N}$ and ${N}_\mathrm{S}$ represent the number of normal and silent speech data, respectively. 
The predicted distributions of normal and silent speech 
 approach to the distribution $M$
 by minimizing the following sum of KL divergences:
\begin{align}
    \mathcal{L}_\mathrm{WKL} &= \mathcal{D}_\mathrm{KL}(M||Q_\mathrm{N}) + \mathcal{D}_\mathrm{KL}(M||Q_\mathrm{S}).
    \label{equation:eq6}
\end{align}
Note that $M$ becomes equal to $P_\mathrm{N}$ 
 if $N_\mathrm{N} / N_\mathrm{S} \rightarrow \infty$,
 which represents the situation of the current VSR for silent speech.

\subsection{Metric learning for within-speech type}
Even when people pronounce the same word in the same speech type,
differences in speaker~\cite{Wand2017ImprovingSL} 
and location~\cite{imko2016HyperarticulationIL} can cause variations in their lip movements.
Inspired by this, we consider within-class variances in viseme probability distributions. 
By reducing such differences through metric learning within each speech type,
the model can classify the distributions into their true visemes more accurately.
To achieve this, 
 we bring all distributions close to a representative distribution if they show the same viseme
 through minimizing the KL divergence $\mathcal{L}_\mathrm{NKL}$ ($\mathcal{L}_\mathrm{SKL}$) for normal (silent) speech data as follows:
\begin{align}
    \mathcal{L}_\mathrm{NKL} &= \mathcal{D}_\mathrm{KL}(P_\mathrm{N}||Q_\mathrm{N}) = {P_\mathrm{N}}\log\dfrac{P_\mathrm{N}}{Q_\mathrm{N}},
    \label{equation:eq1}\\
    \mathcal{L}_\mathrm{SKL} &= \mathcal{D}_\mathrm{KL}(P_\mathrm{S}||Q_\mathrm{S}) = {P_\mathrm{S}}\log\dfrac{P_\mathrm{S}}{Q_\mathrm{S}}.
    \label{equation:eq2}
\end{align}


\section{Evaluation}
This section describes experiments conducted to evaluate 
 the performance of the proposed method for silent speech VSR.

\subsection{Database}
To conduct our experiments, we used the AV Digits database~\cite{Petridis2018VisualOnlyRO}, the only publicly available dataset that contains both normal and silent speech.
This limited dataset comprises 39 speakers who repeated 10 brief phrases (including ``Excuse me'', ``Goodbye'', ``Hello'', ``How are you'', ``Nice to meet you'', ``See you'', ``I am sorry'', ``Thank you'', ``Have a good time'', ``You are welcome'') five times each in both speaking modes.
Following the methodology of~\cite{Petridis2018VisualOnlyRO}, we split the dataset into training, validation, and test sets of 1,000 utterances (from 20 speakers), 400 utterances (from 8 speakers), and 550 utterances (from 11 speakers), respectively.
Additionally, we augmented the amount of training data for normal speech by incorporating the OuluVS2 database~\cite{Anina2015OuluVS2AM}, which includes 1560 utterances (from 52 speakers) with the same vocabulary as the AV Digits database.
In total, we utilized 2560 utterances (from 72 speakers) for normal speech and 1,000 utterances (from 20 speakers) for silent speech during the training phase by combining these two datasets.
Since the datasets only include text-based labels, we transformed them into viseme sequences by first translating texts to phoneme sequences using the Carnegie Mellon Pronouncing (CMU) Dictionary, and then mapping the phonemes to visemes according to Lee and Yook’s approach~\cite{Lee2002AudiotoVisualCU}.

\subsection{Training settings}
Following previous studies~\cite{Fenghour2020LipRS, Fenghour2019DecoderEncoderLF, Fenghour2021AnEC}, 
the visual and language models were trained separately while fixing the visual front-end with weights pre-trained by~\cite{Afouras2018DeepAS}. 
Note that we conducted a thorough exploration of various parameters, including the weights for loss terms, and adopted the parameters that achieved the best performance. 
To train the visual model, we used the Adam optimizer~\cite{Kingma2014AdamAM} with an initial learning rate of $10^{-5}$ and a mini-batch size of 64 for 50 epochs, and decreased the learning rate by 50\% if the validation loss did not improve over five epochs. For the language model, we employed the same settings, except for the initial learning rate and mini-batch size, which were set to $5 \cdot 10^{-4}$ and 10, respectively. During all experiments, the language model was trained on the viseme sequences predicted by the visual model using normal speech data.

{\tabcolsep=1.3mm
\begin{table}[t]
  \caption{
  Performance comparison of different loss functions during training. 
  A lower score indicates higher accuracy.
  }
  \vspace{-2.5mm}
  \label{tab:lossfunction}
  \centering
  \begin{tabular}{lrrrr} 
  \toprule
    \multirow{3}{*}{Loss function} & \multicolumn{2}{c}{VER (\%)} & \multicolumn{2}{c}{WER (\%)} \\
    \cmidrule(lr){2-3} \cmidrule(lr){4-5}
     & Normal & Silent & Normal & Silent \\
    \midrule
    $\mathcal{L}_\mathrm{NCE}$ & 5.87 & 12.49 & 11.58 & 22.55 \\
    \midrule    
    $\mathcal{L}_\mathrm{NCE}$ + $\mathcal{L}_\mathrm{SCE}$ (Baseline) & 5.30 & 7.48 & 10.21 & 12.91 \\
    \quad + $\mathcal{L}_\mathrm{NKL}$ & 5.24 & 7.77 & 7.36 & 13.45 \\
    \quad + $\mathcal{L}_\mathrm{SKL}$ & 5.10 & 7.20 & 8.70 & 12.42 \\
    \quad + $\mathcal{L}_\mathrm{NKL}$ + $\mathcal{L}_\mathrm{SKL}$ & 5.22 & 7.07 & 8.48 & 12.24\\
    \midrule
    $\mathcal{L}_\mathrm{NCE}$ + $\mathcal{L}_\mathrm{SCE}$ + $\mathcal{L}_\mathrm{KL}$ & 5.21 & 7.17 & 8.12 & 11.06 \\
    \quad + $\mathcal{L}_\mathrm{NKL}$ & 5.15 & 6.92 & 7.85 & 10.85 \\
    \quad + $\mathcal{L}_\mathrm{SKL}$ & 5.13 & 7.35 & 9.36 & 10.71 \\
    \quad + $\mathcal{L}_\mathrm{NKL}$ + $\mathcal{L}_\mathrm{SKL}$ & 5.11 & 6.96 & 7.85 & 10.73 \\
    \midrule
    $\mathcal{L}_\mathrm{NCE}$ + $\mathcal{L}_\mathrm{SCE}$ + $\mathcal{L}_\mathrm{WKL}$ & 5.16 & 6.89 & 8.70 & 10.73 \\
    \quad + $\mathcal{L}_\mathrm{NKL}$ & 5.09 & 6.71 & 8.00 & 10.88\\
    \quad + $\mathcal{L}_\mathrm{SKL}$ & 5.10 & 6.89 & 8.76 & 11.00 \\
    \quad + $\mathcal{L}_\mathrm{NKL}$ + $\mathcal{L}_\mathrm{SKL}$ & 5.03 & \textbf{6.66} & 8.20 & \textbf{9.97} \\    
    \bottomrule
  \end{tabular}
  \vspace{-2mm}
\end{table}
}

\subsection{Evaluation metrics}
For evaluation, we used two metrics: viseme error rate (VER) for the visual model outputs and word error rate (WER) for the language model outputs. 
We calculated VER as follows:
\begin{align}
    \mathrm{VER} = \frac{V_{S}+V_{D}+V_{I}}{V_{N}},
    \label{equation:eq7}
\end{align}
where $V_{N}$ is the total number of visemes in the ground truth. $V_{S}$, $V_{I}$, and $V_{D}$ represent the number of visemes substituted for wrong classifications, inserted for those not to be present, and deleted for those to be present, respectively.
WER refers to the same concept as VER 
yet is against words instead of visemes.

\subsection{Experimental results}
We conducted 13 experiments with different combinations of loss functions to evaluate the accuracy of our proposed method, and the results are summarized in Table \ref{tab:lossfunction}.
Initially, to confirm the issue with silent speech, we trained the model solely on normal speech data using the cross-entropy $\mathcal{L}_\mathrm{NCE}$.
This resulted in the largest performance drop for silent speech in both VER and WER, with values of 6.62\% and 10.97\%, respectively, compared to normal speech.
To serve as a baseline for comparison, we used the model trained only with the classification loss function $\mathcal{L}_\mathrm{NCE} + \mathcal{L}_\mathrm{SCE}$, which achieved higher accuracy than the adaptation approach 
pre-trained with normal speech and then fine-tuned with silent speech.
Additionally, we incorporated two kinds of metric learning for inter-speech type, namely $\mathcal{L}_\mathrm{KL}$ and $\mathcal{L}_\mathrm{WKL}$.
Both approaches improved accuracy for silent speech, leading to a 1.85\% and 2.18\% decrease in WER, respectively, compared to the baseline.
These results demonstrate the effectiveness of using metric learning between normal and silent speech, and treating them with equal consideration.
For further improvement, we added metric learning for within-speech type, i.e., using only $\mathcal{L}_\mathrm{NKL}$, only $\mathcal{L}_\mathrm{SKL}$, or both.
The best accuracy achieved for silent speech was 6.66\% in VER and 9.97\% in WER, using the loss function $\mathcal{L}_\mathrm{NCE} + \mathcal{L}_\mathrm{SCE} + \mathcal{L}_\mathrm{WKL} + \mathcal{L}_\mathrm{NKL} + \mathcal{L}_\mathrm{SKL}$, which showed that metric learning within each speech type could reduce the variances in the output distributions.

\begin{figure}[t]
  \centering
  \vspace{-6mm}
  \includegraphics[width=0.93\linewidth]{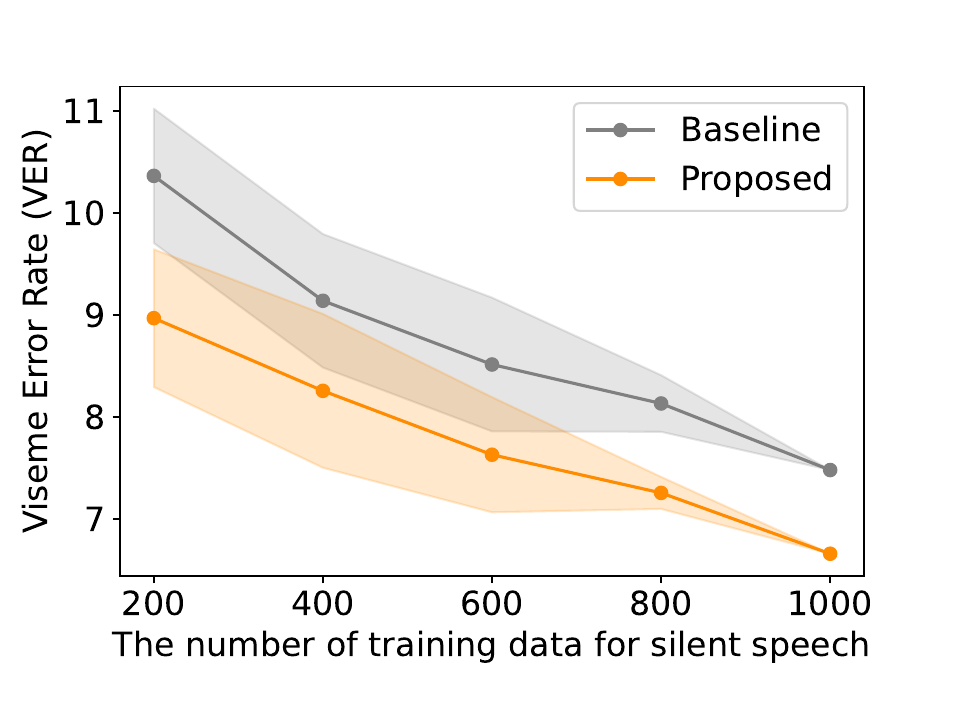}
  \vspace{-5mm}
  \caption{
  Impact of decreasing training data for silent speech on VER, from 1,000 to 200 by 200.
  The number of normal speech data remained constant at 1,000. 
  The shaded area represents the range $m\pm\sigma$, 
  where $m$ is the mean and $\sigma$ is the standard deviation of VER over five experiments.}
  \label{fig:fewerdata}
  \vspace{-3mm}
\end{figure}

In these experiments, we utilized 2,560 normal speech data and 1,000 silent speech data in the training process.
However, recent major studies~\cite{Afouras2018DeepAS, Afouras2018LRS3TEDAL, Cooke2006AnAC} have employed VSR datasets consisting of tens of thousands of normal speech data, emphasizing the need to evaluate the effectiveness of our proposed method in more realistic scenarios.
To assess its robustness to an imbalanced amount of data used for each speech type, we reduced the number of utterances per speaker from 50 to 40, 30, 20, and 10 in silent speech data, by modifying ${N}_\mathrm{S}$ in Equation \ref{equation:eq5}.
Each experiment using the same number of data was conducted five times, with different training data excluded in each run.
Figure \ref{fig:fewerdata} shows that the proposed method consistently achieved a lower VER for silent speech than the baseline method across all training data settings.
Notably, the proposed model trained on 400 silent speech data exhibited comparable accuracy to the baseline model trained on twice the amount of data.

\section{Conclusion}
We proposed a novel approach to improve the accuracy of silent speech prediction 
 using a VSR model trained on normal speech. 
Specifically, we presented a viseme-based metric learning method 
 that addressed the limited availability of silent speech datasets 
 by focusing on the common viseme representations between normal and silent speech. 
Our approach employed metric learning techniques 
 to learn visemes inter- and within-speech types, 
 allowing us to efficiently leverage silent speech data 
 while accounting for variations within the same speech type. 
Through extensive experimentation with various loss functions, 
 we demonstrated that our proposed method consistently 
 outperformed the baseline approach, even under challenging conditions.
In future work, we plan to extend our approach to larger normal speech datasets to facilitate VSR on silent speech data with unknown vocabulary, which was not included in the 10 phrases used in our experiments.


\section{Acknowledgements}
This work is supported in part by JSPS KAKENHI Nos. 21H05054, 22J22424, 22KJ2959.



\bibliographystyle{IEEEtran}
\bibliography{mybib}

\end{document}